\documentstyle[11pt,aaspp4]{article}

\begin{document}

\title{Cosmological Magnetic Fields Limits in an Inhomogeneous
Universe}

\author{Pasquale Blasi, Scott Burles, and Angela V. Olinto}
\affil{Department of Astronomy and Astrophysics, and Enrico
Fermi Institute, \\ University of Chicago, 5640 S. Ellis
Ave, Chicago, IL 60637}

\authoremail{blasi@oddjob.uchicago.edu,
scott@spectro.uchicago.edu, and olinto@oddjob.uchicago.edu}

\begin{abstract}

We study the effect of inhomogeneities in the matter
distribution of the universe on the Faraday rotation of
light from distant  QSOs and derive new limits on
the cosmological magnetic field. The matter distribution in the Universe
is far from being homogeneous and, for the redshifts of interest to
rotation measures (RM), it is well described by the observed  Ly-$\alpha$
forest.  We use a log-normal distribution to model the Ly-$\alpha$
forest and assume that a cosmological magnetic field is frozen into the
plasma and is therefore a function of the density inhomogeneities. 
The Ly-$\alpha$ forest results are much less sensitive to the
cosmological magnetic field coherence length than those for a
homogeneous universe and show  an increase in the magnitude of the
expected RM for a given field by over an order of magnitude. The forest
also introduces a large scatter in RM for different lines-of-sight with a
highly non-gaussian tail that renders the variance and the mean RM
impractical for setting limits. The
$median|{\rm RM}|$ is a better statistical indicator which we use to
derive the following limits using the observed RM for QSOs between $z =
0$ and $z = 2.5$.  We set $\Omega_b h^2 = 0.02$ and get for
cosmological fields coherent accross the present horizon, $B_{H_0^{-1}}
\la  10^{-9}$ G in the case of a Ly-$\alpha$ forest which is
stronger than the limit for a homogeneous universe, $B_{H_0^{-1}}^h
\la 2 \times 10^{-8}$ G; while for  50
Mpc coherence length, the inhomogeneous case gives $B_{50 {\rm Mpc}} \la
6 \times 10^{-9}$ G while the homogeneous limit is
$B_{50 {\rm Mpc}}^h \la  10^{-7}$ G and 
for coherence length equal to the Jeans length,
$B_{\lambda_J} \la  10^{-8}$ G for the Ly-$\alpha$ case while 
$B_{\lambda_J}^h \la  10^{-6}$ G.

\end{abstract}

\section{Introduction}

The nature and structure of extragalactic magnetic fields are
currently unknown. Magnetic fields in clusters of galaxies and in bridges
between clusters have been observed (Kim et al. 1989; Vallee
1990;  Kronberg 1994 and references therein) but  fields
in cosmological scales are yet to be established. Cosmological fields 
produced before galaxy formation  would be present throughtout the
Universe today and could be detected by Faraday rotation observations. A
number of models for the generation of such primordial fields have been
suggested (e.g., Harrison 1969, 1970, 1973; Hogan 1983; Turner \& Widrow
1988; Vachaspati 1991;  Quashnock, Loeb, \& Spergel 1989; Ratra 1992;
Enqvist \& Olesen 1993; Dolgov \& Silk 1993;  Cheng
\& Olinto 1994; Kibble
\& Vilenkin 1995; Gasperini, Giovannini, \& Veneziano 199, Lemoine
\& Lemoine 1995; Baym, Bodecker,
\& McLerran 1996;  Sigl, Olinto,
\& Jedamzik 1997; Joyce \& Shaposhnikov 1997; Giovannini \& Shaposhnikov
1997) with hopes of seeding magnetic fields in galaxies. The evolution
of these fields has been studied before (Jedamzik, Katalinic,
\& Olinto 1998; Brandenburg, Enqvist, \& Olesen 1997)  and around
recombination (Wasserman 1978; Kim, Olinto,
\& Rosner 1996) but theoretical expectations for the  present structure
of these fields has not been contrasted with observations.  Here we take
a first step in this direction by considering the effect of densities
inhomogeneities in the  present Universe on Faraday rotation measures
and find much stronger limits   on the primordial field than the
previously assumed case of a homogeneous density distribution.

As early
researchers realized (Rees \& Reinhardt 1972; Nelson 1973; Kronberg
\& Simard-Normandin 1976), the most effective method for studying large
scale magnetic fields is the detection of Faraday rotation from
extragalactic sources. 
 Faraday rotation occurs when polarized
electromagnetic radiation travels through a magnetized medium.
The polarization vector rotates an angle, $\psi$,  proportional to the
magnetic field along the line-of-sight, $B_\parallel(l)$,  and varies
with the observed wavelength as  $\lambda_{obs}^{-2}$.  The rotation
measure is defined as, 
\begin{equation}
 RM \equiv {\psi \over \lambda_{obs}^2} = {e^3 \over {2
\pi m_e^2 c^4}} 
\int_0^{l_s} n_e(l) B_{\parallel}(l) \left({\lambda(l) \over
\lambda_{obs}}
\right)^2 dl \ \ , 
\end{equation}
where  $n_e$ is the number density of electrons
(with charge $e$ and mass $m_e$),
$\lambda(l)$ is the wavelength at a physical position $l$ along the
line-of-sight to a source at physical distance $l_s$, and $c$ is the
speed of light. 

This method has been quite sucessful in determining the structure of
the Galactic magnetic field by observing both RM and the dispersion
measure of pulsar  radio  emission. With the dispersion
measure data, the electron number density can be estimated and a
$B_\parallel$ determined.  For extragalactic sources, no
dispersion measure data is attainable and the electron density has to
be modeled theoretically.

Prior studies of extragalactic RM have generally used  oversimplified
models for the electron density distribution given what we
presently know about the matter distribution in the Universe. In fact,
the electron density is usually assumed to be a homogeneous distribution
that only scales with redshift,
$z$,  as
$n_e \propto (1+z)^{3}$. Instead, it has been clear for a number of
years that the matter in the intergalactic medium is not
homogeneously distributed.
A new picture of the intergalactic medium has emerged in which
the baryon fluctuations trace the dark matter fluctuations on scales
larger than the Jeans length  (Cen et al. 1994).
This picture fits well with high resolution spectroscopic observations of
the ubiquitous high-redshift neutral
hydrogen absorption clouds known as the Lyman-$\alpha$ forest (Kirkman
\& Tylter 1997, Hu et al. 1995, Lu et al. 1996)
and with N-body+hydro simulations of large-scale structure
formation (Rauch et al. 1997, Zhang et al. 1998)

In this Letter, we study the RM  of distant QSOs
assuming that the electron distribution follows the observed
Lyman-$\alpha$ forest distribution. We show that not only the variance
and mean RM for different lines-of-sight can differ by more than an order
of magnitude from  the case of a homogeneous universe but
their scatter is too large to determine a significant limit. The
dependence on the cosmological magnetic field coherence scale is
almost washed out becoming  subdominant to the density
fluctuations scale.  To determine a limit, we use instead the
$median|{\rm RM}|$  and   find that for  $\Omega_b
h^2 = 0.02$ and $\Omega_M=1$ the cosmological field is $B
\la  10^{-9} - 10^{-8}$ G for coherence lengths varying from the
Jeans length to the horizon scale, $H_0^{-1}$. 

In fact,  for cosmological fields coherent accross the present horizon,
the case of a Lyman-$\alpha$ forest gives
$B_{H_0^{-1}}
\la  10^{-9}$ G which is
stronger than the limit for a homogeneous universe, $B_{H_0^{-1}}^h
\la 2 \times 10^{-8}$ G; while for  50
Mpc coherence length, the inhomogeneous case gives $B_{50 {\rm Mpc}} \la
6 \times 10^{-9}$ G while the homogeneous limit is
$B_{50 {\rm Mpc}}^h \la  10^{-7}$ G and 
for coherence length equal to the Jeans length,
$B_{\lambda_J} \la  10^{-8}$ G for the Lyman-$\alpha$ case while 
$B_{\lambda_J}^h \la  10^{-6}$ G.

\section{The Electron Density Distribution}

For distant extragalactic sources, the integral in Eq. (1) can
be written as an integral over redshift using the radial
displacement-redshift relation:
\begin{equation}
 dl(z) = - {c \ dz \over H_0 (1+z) \sqrt{ \Omega_M
(1+z)^3 +
\Omega_\Lambda + (1- \Omega_M - \Omega_\Lambda) (1+z)^2} } 
\end{equation}
where $dl$ is the proper displacement at redshift $z$, $H_0$ is the
present Hubble parameter,
$\Omega_M$ is the ratio of the matter density to the critical density
in the Universe, and 
$\Omega_\Lambda$ is the vacuum energy density over the critical density
(see, e.g., Peebles 1993). Following the usual choices for
the cosmological parameters,  we assume $h = 0.65$,
$\Omega_M =1$, and $\Omega_\Lambda =0$ below. (We have also
considered the case of $\Omega_M =0.2$ and $\Omega_\Lambda =0.8$, which
changes the result only by a factor  $\la 2$.)

The electron density along the line-of-sight to a QSO depends on
the mass distribution, the baryon fraction,  and the
ionization fraction. 
Observations of the Lyman-$\alpha$ forest in
QSO spectra give evidence of  the neutral hydrogen distribution
along the line-of-sight which we assume to trace the bulk of the baryon
distribution. Given a value of the baryon number
density over the critical,
$\Omega_b$, we can model the electron
distribution by assuming an ionization fraction. We assume that the
bulk of the baryons is ionized hydrogen and helium (24\%) with an
average density given by $\Omega_b$ and that the spatial dependence of
the baryon distribution (hence, the electron distribution) is the same
as the neutral hydrogen distribution.

In modelling the intergalactic
medium, we use the analytical approximation
of a  log-normal density distribution  for the Lyman-$\alpha$
forest (Bi \& Davidsen 1997, and Coles
\& Jones 1991) which gives
a simple and consistent description of the 
observations. The log-normal distribution sets the baryon density
fluctuations at the 
Jeans length for photoionized clouds, $\lambda_J$,  which can be
written in comoving coordinates as (Gnedin \& Hui 1996, 
Bi \& Davidsen 1997,  and Bond, Szalay, \& Silk, 1988):
\begin{equation}
 {{2 \pi} \over {\lambda_J}} = 7.4 \sqrt{\Omega_0(1+z) {{10^4 K} 
        \over {T}}} \, h \, \rm{Mpc}^{-1} \ .
\end{equation}

The log-normal distribution gives the electron density
distribution as a function of one parameter, $\sigma(z)^2$, which
is the linear variance over the Jeans length.
For standard CDM,  a good fit to $\sigma(z)$  can be written as
\begin{equation}
 \sigma(z) = 0.08096 + 5.3869 (1+z)^{-1} - 4.21123
        (1+z)^{-2} + 1.4433 (1+z)^{-3} \ , 
\end{equation}
while $\Lambda$CDM  gives $\sigma(z)$  within 20\% of standard CDM (Bi
\& Davidsen, 1997).

We  simulate 10,000 random lines-of-sight to high-redshift QSOs
by randomly drawing electron densities for cells given by the
Jeans length. For each Jeans length cell, 
we calculate the electron over-density ($\delta$) with the 
log-normal equation:
\begin{equation}
P(\delta) d\delta = {1\over {\sqrt{2\pi} \sigma (1+\delta)}} {\rm \,
exp} \left({-[\,{\rm ln}(1+\delta) + \sigma^2/2]^2\over
{2\sigma^2}}\right) \, d\delta \ .
\label{Pdelta}
\end{equation}
The physical electron density is then given in terms of the
over-density calculated above,
\begin{equation}
  n_e (z) = {\bar n}_e (z)  (1+\delta) \ ,
\end{equation}
where the mean electron density as a function of redshift, ${\bar n}_e(z)
= {\bar n}_e(0)  (1+z)^3$, and ${\bar n}_e(0) = 9.1 \times 10^{-6} {\rm
cm}^{-3}
\Omega_b \, h^2$. We set $\Omega_b \, h^2 = 0.02$ from D/H
measurements in high-redshift QSO systems (Burles \& Tytler 1998a,
1998b) together with the predictions of big bang nucleosynthesis
(see, e.g., Copi, Schramm,
\& Turner 1995). This gives ${\bar n}_e(0) = 1.8\times 10^{-7}  {\rm
cm}^{-3} $ independent of
$H_0$.

We add the rotation measures contributed by each Jeans length along
the line-of-sight from the QSO redshift to z=0.  
In each cell,   the rotation
measure,
\begin{equation}   
\Delta \, RM = 8.1 \times 10^5 \left({\rm rad \over m^2}\right) \,
{1 \over (1+z)^{2}} \; \left({n_e(z) \over {\rm cm}^{-3}}\right) \;
\left({{B}_{||} (z)
\over \mu{\rm G}}\right)\;  \left({\Delta l \over {\rm Mpc}}\right) \ \
,
\end{equation}  
where $\Delta l = \lambda_J / (1+z)$ and can be related to $\Delta z$
through Eq. (2).

We assume that the large scale magnetic field we are constraining 
predates the formation of structure and is considered frozen into the
plasma. The conductivity of the plasma in the Universe is large enough
for the  frozen in assumption to hold on scales $\ga$  1 A.U.
(Cheng \& Olinto, 1995) which is well below the scales of interest here.
The field direction is then taken to be fixed between different magnetic
field coherence lengths, $L_c$,  while the field magnitude changes as a
function of the electron density. For simplicity, we assume that the
overdensities are spherical on average and thus scale the field
magnitude as $B \propto n_e^{2/3}$. 
 
Therefore, the magnetic field projected along the line-of-sight,
${B}_{||} (z) $, is  fixed within each cell to be 
${B}_{||} (z) = B (z) \; cos \theta =
B_0 (n_e /{\bar n}_e(0))^{2/3} \; cos \theta$, where $B_0$ is
the magnitude of the cosmological field  in comoving
coordinates (actual value at $z=0$) and $\theta$ is the angle between
the line-of-sight and the direction of the cosmological field.
$\theta$ is chosen randomly   for  each magnetic field coherence length
scale. For magnetic field coherence length, $L_c =
\lambda_J/(1+z) $, the random angle of the field is chosen at each step,
while for $L_c = N
\lambda_J$, the random direction is changed after $N$ steps.

The coherence length for a given cosmological field depends on the
particular magnetogenesis model and the evolution of the field up to
recombination. Given the damping of the short wavelength modes up to
the Silk scale (Jedamzik, Katalinic, \& Olinto 1998) the coherence scale
is likely to be above $\lambda_J$, which today is $\lambda_J \simeq 1$
Mpc. We chose to study three choices of $L_c$ that span the range of
interest for plausible cosmological fields: $\lambda_J$, 50 Mpc, and
$H_0^{-1}$.   As discussed below, our results are not very sensitive to
the coherence length of the field and our conclusions can be applied to
any model of  magnetogenesis where $L_c \ga \lambda_J$.

\section{Results}

After integrating Eq. (1) for 10,000 lines-of-sight per 0.1 redshift
interval, we display our results in Figs. 1 and 2, where we plot the
median of the absolute value of the RM ($median |{\rm RM}|$, circles in
Fig. 1), the mean absolute value of RM ($<|{\rm RM}|>$, squares in Fig.
1), and the variance ($<{\rm RM}^2>$, triangles in Fig. 2). For
comparison, we also show the same functions for the case of a
homogeneous electron distribution throughout the universe ($median |{\rm
RM}|$ are the solid lines in Fig. 1, the 
$<|{\rm RM}|>$ are the dashed lines in  Fig. 1, and the variance $<{\rm
RM}^2>$ is the solid line in Fig. 2). We also bin the available data on
QSO rotation measures in two redshift bins, $z = 0 - 1.5$ and $z=1.5 -
2.5$ and display the $<|{\rm RM}|>$ (dashed-dotted line)  and the $median
|{\rm RM}|$ (dotted line) in Fig. 1.

Not surprisingly we find that the Lyman-$\alpha$ forest introduces a
very large scatter in the RM for different lines-of-sight. Not only
 the magnitude of the variance is many
orders of magnitude greater than the case of a homogeneous universe
but the variance and the mean RM vary significantly within our
unrealistically large simulated  sample of 10,000 lines-of-sight per
$\Delta z = 0.1$ redshift interval 
(Fig. 2). With such a large variance, the limited data set available at
present is not sufficient to constrain well the cosmological magnetic
field.  In fact, Oren \& Wolfe (1995) showed how sensitive is the
redshift dependence of the variance to different binning procedures and
how the observed distribution is non-gaussian. The trends in the
variance as a function of redshift in Fig. 10 of Oren \& Wolfe (1995) 
(where they plot different binning of data compiled by Welter, Perry, \&
Kronberg 1984) are incompatible with the variance behaviour in a
homogeneous universe (see Fig. 2). In particular, the homogeneous case
has a monotonically increasing variance with redshift while the
inhomogeneous case can reproduce the behaviour shown in their Fig. 10.
In this case, the non-gaussian tail of the RM is due to a few large
density clouds along  random lines-of-sight. The non-gaussian tail can
also be due to non cosmological magnetic fields such as small scale
structure unaccounted for in the Galactic magnetic field.

Of the statistical indicators we studied, the variance is the noisiest
followed by the mean and the median is the most stable indicator.   This
behaviour is evident in Figs. 1 and 2. The fact that the mean and the
mediam differ significantly in the case of an inhomogeneous universe is
further evidence of the non-gaussian behaviour of the rotation
measures. The mild magnetic coherence length dependence of the median
and the mean, is such that as the coherence length grows the two
indicators approach each other. We use the best statistical indicator,
the  $median |{\rm RM}|$, to  set our limits for both the homogeneous and
inhomogeneous cases. We only set limits versus detections due to the
presently unsurmountable challenge  of separating locally generated
contributions to the rotation measure from cosmologically generated ones.

The available date on QSO RMs is sparse and dominated by the Galactic
magnetic field contribution. Courageous attempts at modelling
and subtracting  the Galactic contribution can be found in
Simard-Normandin
\& Kronberg 1980, Simard-Normandin, Kronberg, \& Burton 1981,
Kronberg \& Perry 19982, Welter, Perry, \& Kronberg 1984, Watson \& Perry
1991, Vall\'ee 1990, 1993, and Oren \& Wolfe 1995. (For a critical
comparison of these attempts see Oren \& Wolfe 1995.) Given the
sparseness of the available data, we chose to bin the Welter, Perry, \&
Kronberg 1984 catalog in two redshift bins ($0 <z < 1.5$ and $1.5 < z <
2.5$) and calculated the mean absolute RM (dashed-dotted line) and the
median absolute RM (dotted line) for this data and compared these with
our predictions.

For the
$\Omega_M=1$ flat universe (with
$\Omega_b h^2 = 0.02$) we get for cosmological fields coherent accross
the present horizon,
$B_{H_0^{-1}}
\la  10^{-9}$ G in the case of a Lyman-$\alpha$ forest which is
stronger than the limit for a homogeneous universe, $B_{H_0^{-1}}^h
\la 2 \times 10^{-8}$ G; while for  50
Mpc coherence length, the inhomogeneous case gives $B_{50 {\rm Mpc}} \la
6 \times 10^{-9}$ G while the homogeneous limit is
$B_{50 {\rm Mpc}}^h \la  10^{-7}$ G and 
for coherence length equal to the Jeans length,
$B_{\lambda_J} \la  10^{-8}$ G for the Lyman-$\alpha$ case while 
$B_{\lambda_J}^h \la  10^{-6}$ G.
 Changing cosmological
parameters to, for example, $\Omega_M =0.2$,$\Omega_\Lambda =0.8$ in a
flat universe, the results only change by a factor of 1.6; we find  $
median \, |RM|$ ($\Omega_M =0.2$,$\Omega_\Lambda =0.8$) $\approx 1.6
\times  median \,  |RM|$ ($\Omega_M=1$,$\Omega_\Lambda =0$), and the
limits on cosmological magnetic fields is 1.6 times lower.
These limits are slightly stronger than limits attained by alternative
methods. For instance,  the limit due to the isotropy of the
cosmic background  radiation is 
$B_{H_0^{-1}} \la  4.4 10^{-9}$ G for $ \Omega_0 = 1$ and $h = 0.65$
(Barrow, Ferreira, Silk 1997). Given the non-gaussian nature
of the RM distribution, the sparseness of the data, and the difficult
Galactic subtraction, it is encouraging to find the RM limit for the
largest coherence length to be consistent with and comparable to the
cosmic background radiation limit.

Our results imply stronger constraints on scenarios of primordial
magnetic field generation and evolution by limiting the spectrum of
primordial fields today both the magnitude and the scale dependence,
i.e., we limit  
$B_0 (L_c) $. This lowers the available primordial seed for
galactic dynamos and for structure formation by primordial fields.  
Our results also help the prospective of detectors pointing back to
ultra-high energy cosmic ray (UHECR) sources, by limiting the
primordial component of the intergalactic medium field. Magnetic effects
on UHECRs will be more important if post recombination fields, limited to
sites of structure formation,  are much stronger than the fossil fields
from the early universe (e.g., Ryu, Kang, \& Biermann 1998).

\acknowledgements

We thank Lam Hui for very useful discussions
on the Lyman-$\alpha$ forest. P.B. was supported by  a fellowship of the
Istituto Nazionale di Fisica Nucleare at the University of Chicago.
 A.V.O. was supported in part by
the DOE through grant DE-FG0291 ER40606 and the NSF through grant
AST 94-20759 at the University of Chicago.

\newpage

\figcaption[fig1.ps]{
Observed and predicted statistics of QSO rotation
measures for magnetic field coherence lengths of
the Jeans length, 50 Mpc, and H$_0^{-1}$.  The predicted absolute
mean (squares) and
median (circles) rotation measure for a 6 nG field in the
Ly-$\alpha$ forest approximation
drawn from 10000 lines of sight at each redshift.
The continuous lines represent the predicted absolute
mean (solid line) and median (dashed) for a homogeneous
electron density and a 6 nG field.
The two bins ($0 <z < 1.5$ and $1.5 < z < 2.5$) show the observed
absolute mean RM (dashed-dotted) and median (dotted) in the
dataset of WPK.
}

\figcaption[fig2.ps]{
The variance of QSO rotation measures predicted
for the Ly-$\alpha$ forest (triangles) and a
homogenenous density (solid line) for a 6 nG
field and the coherence lengths shown in Figure 1.}


\begin{references}


Barrow, J.D.,  Ferreira, P.G., \& Silk, J.,  1997, Phys. Rev. Letters,
78,  3610

Baym, G.,  B\"odecker, D., \& McLerran, L., 1996, Phys.
Rev. D, 53, 662 

 Brandenberger,R. H.,   Davis, A.-C., Matheson, A. M. \& 
Trodden, M., 1992, Phys. Lett.  B, 293, 287 

Bi \& Davidsen 1997, ApJ, 479, 523 

Bond, R., Szalay, A., \& Silk, J., 1988, ApJ, 324, 627

Brandenburg, A., Enqvist, K., \&  Olesen, P. 1997,  Phys.
Lett., B391, 395 

Burles, S.,  \& Tytler, D., 1998a, ApJ, 499, 699; 

Burles, S., \& Tytler, D., 1998b,  ApJ, 507, in press

Cen, R., Miralda-Escude, J., Ostriker, J. P. \& Rauch, M. 1994, ApJ,
437, L9

Cheng, B. \& Olinto, A., 1994, Phys. Rev. D, 50,2421


Coles, P. \& Jones 1991, MNRAS, 248, 1

Copi, C.,  Schramm, D.N.,  \& Turner, M.S.,  1995, Science

Dolgov, A. D. 1993,
 Phys. Rev. D, 48, 2499 

Dolgov, A. D. \&  Silk, J.
1993, Phys. Rev. D, 47, 3144  

Enqvist, K., \& Olensen, P. 1994,  Phys. Lett. B, 329,
195 


Garretson, W. D.,  Field, G. B. \&  Carroll, S. M. 1992, 
Phys. Rev. D, 46, 5346   

Gasperini, M.,  Giovannini, M. \& Veneziano, G. 1995
Phys. Rev. Lett., 75,  3796 

Giovannini \& Shaposhnikov
1997

Gnedin,  \& Hui, L., 1996, ApJ, 472, 73

 Hogan,   C. J. 1983, Phys. Rev. Lett., 51, 1488  

 
Harrison, E. R. 1970, MNRAS, 147, 279

Harrison, E. R. 1973,  MNRAS, 165, 185 

Hu, E. M., Kim, T.-S., Cowie, L. L., Songaila, A. \& Rauch, M. 1995, AJ,
110, 1526

Jedamzik, K.,  Katalinic, V., \&  Olinto, A.V. 1998,
Phys. Rev. D, 57, 3264 

Joyce \& Shaposhnikov 1997; 

 Kibble, T. W. B., \&  Vilenkin, A., 1995, Phys. Rev. D, 52, 679 


Kim, E.,  Olinto,  A. V. \&  Rosner,R., 1996, Ap. J.,
467

Kim,  K.-T.,    Kronberg, P.P., Giovannini, G., \& Venturi, T,  1989,
Nature, 341, 720
  
Kim,  K.-T.,   Tribble,P.C. \&  Kronberg, P.P., 1991,
 ApJ,379, 80 

Kirkman, D. \& Tytler, D. 1997, ApJ, 484, 672

Kronberg, P. P., \& Simard-Normandin, M. 1976, Nature, 263, 653

Kronberg, P.P., \& Perry, J.J.,  1982, ApJ, 263, 518

Kronberg,  P. P., 1994, Rep. Prog. Phys., 57, 325


Kulsrud, R.,   Ryu, D., Cen, R., \&  Ostriker,  J. P. , 1997
ApJ, 480, 481 

Lemoine
\& Lemoine 1995; 

Lu, L., Sargent, W. L. W., Womble, D. S., \& Takada-Hidai, M. 1996, ApJ,
472, 509

Martin, A. P. , \&
 Davis, A.-C. 1995,  Phys. Lett. B, 360, 71  

Nelson, A.H., 1973, Publ. Ast. Soc. Japan, 25, 489

Oren, A.L., and Wolfe, A.M., 1995, ApJ 445, 624

 Quashnock, J.,  Loeb, A., \& Spergel, D.N. 1989, ApJ, 344,
L49


Rauch, M., Miralda-Escude, J., Sargent, W. L. W., Barlow, T. A.,
Weinberg, D. H., Hernquist, L., Katz, N., Cen, R., \& Ostriker, J. P.
1997, ApJ, 489, 7


 
Ratra, B., 1992,  Ap. J, 391, L1 

Ratra, B., 1992, Phys. Rev. D, 45, 1913 


Rees, M.J., \& Reinhardt, M., 1972, A\&A, 19,104

Ryu, D.,  Kang, H., \& Biermann, P., 1998,  A\&A


Sigl, G.,  Jedamzik, K., \& Olinto, A.V. 1997, Phys.
Rev. D, 55, 4582 

Simard-Normandin, M., \& Kronberg, P.P., 1980, ApJ, 242, 74

Simard-Normandin, M., Kronberg, P.P., \& Burton, S.  1981, ApJ Suppl.,
45, 97

 Turner, M. S. \&  Widrow, L. M. 1988, 
Phys. Rev. D, 30, 2743 

Vallee, J. P., 1990, ApJ 360, 1

Vallee,   J. P.,1993, M.N.R.A.S.  264, 665 

Vaschaspati, T. 1991, Phys. Lett. B, 265, 258

Wasserman, I. 1978, Ap. J., 224, 337 


Watson, A.M.,  \& Perry, J.J., 1991, MNRAS, 248,58

Welter, G.L.,  Perry, J.J.,  \& Kronberg, P.P., 1984, ApJ 279, 19

  
Zhang, Y., Meiksen, A., Anninos, P., Norman, M. L. 1998, ApJ, 495, 63


\end{references}
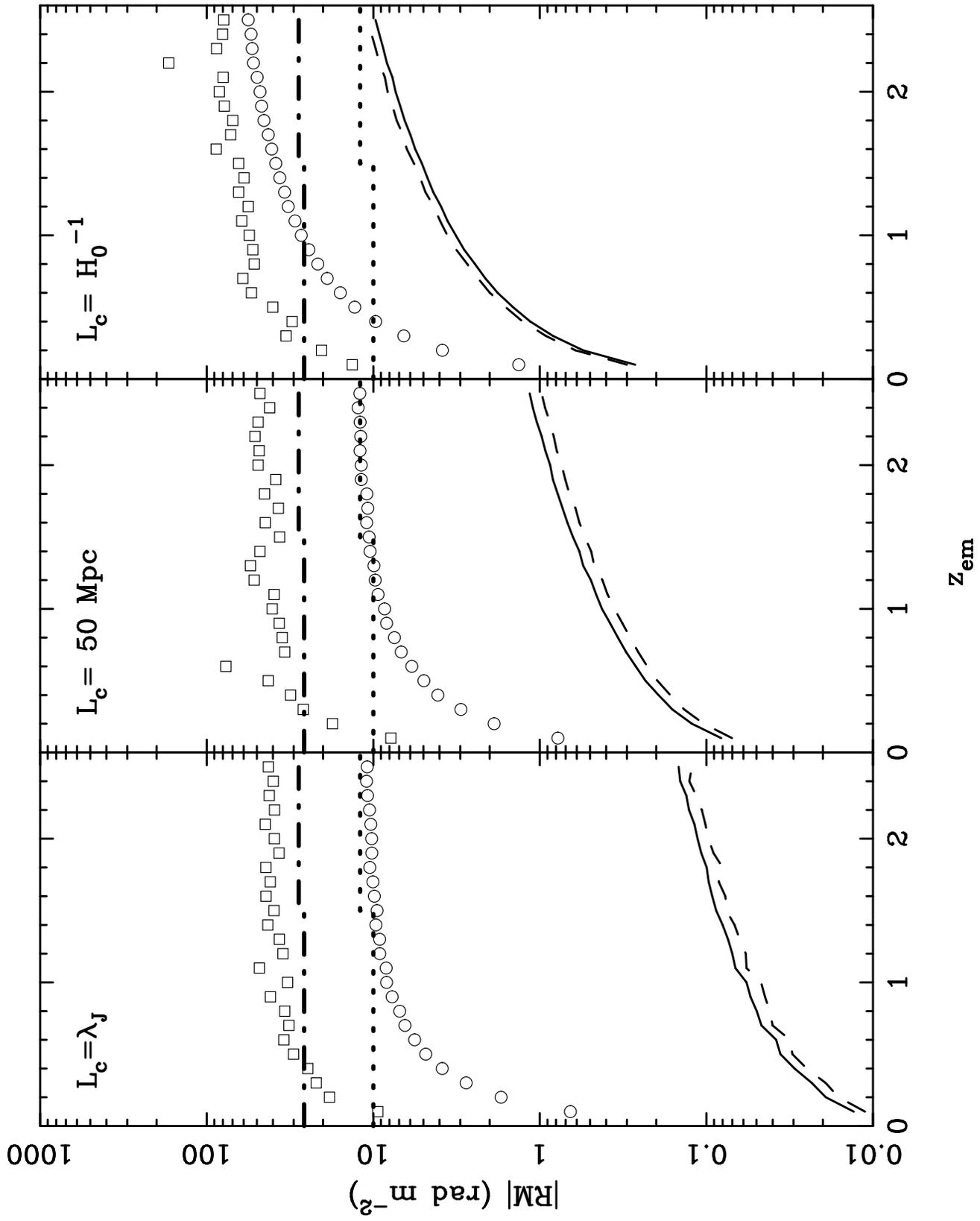
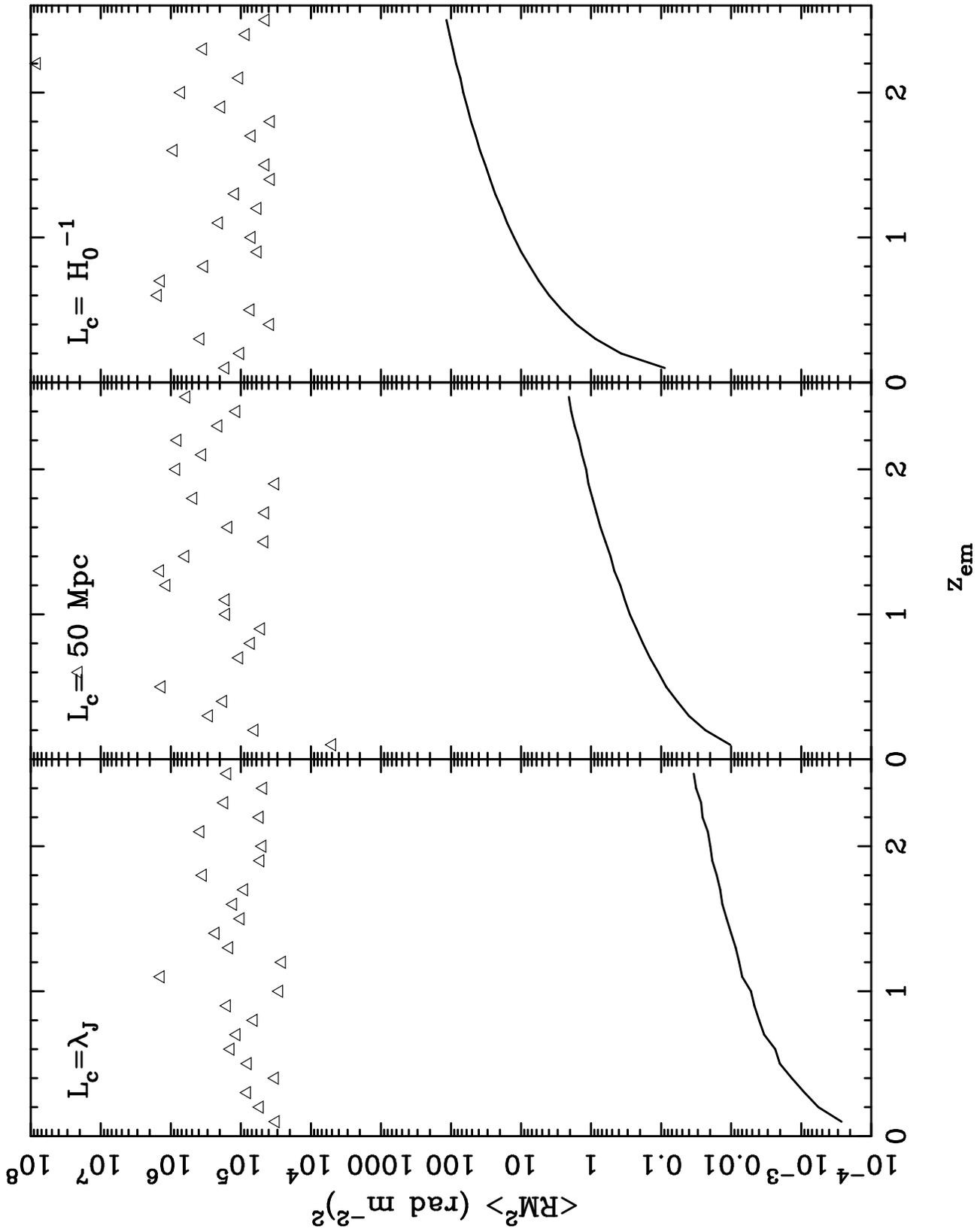
\end{document}